\documentclass[12pt]{article}
\begin{document}
\title{Stochastic Quantization of the Time-Dependent Harmonic Oscillator}
\author{
F.~Haas\footnote{ferhaas@exatas.unisinos.br}\\
Ci\^encias Exatas e Tecnol\'ogicas -- UNISINOS \\
Av. Unisinos, 950\\
93022-000 S\~ao Leopoldo, RS, Brazil
}
\maketitle
\begin{abstract}
\noindent We extend the stochastic quantization method recently developed by Haba and Kleinert to non-autonomous mechanical systems, in the case of the time-dependent harmonic oscillator. In comparison with the autonomous case, the quantization procedure involves the solution of a nonlinear, auxiliary equation. 
\end{abstract}

\hspace{.8cm} {\it PACS numbers: 03.65.-w; 02.50.Ey; 03.65.Ca}

\hspace{.8cm} {\it Keywords: stochastic quantization; time-dependent harmonic \\ \strut \hspace{3.45cm} oscillator; Schr\"odinger equation.}

\vskip 1cm
\noindent Corresponding author: F. Haas\\
Postal address: Unisinos, Ci\^encias Exatas e Tecnol\'ogicas, Av. Unisinos, 950, \strut \hspace{2.8cm} 93022-000, S\~ao Leopoldo, RS, Brazil\\
Telephone: 55 51 590 1655\\
Fax: 55 51 590 8172

\newpage
\section{Introduction}
The fundamental goal of stochastic quantization \cite{pena}-\cite{olavo} is to reproduce the whole content of quantum mechanics by means of classical equations subjected to random perturbations. Presently, there remains the controversy about the physical origin of the noise, which is supposed to arise, for instance, from a fluctuating metrics or from fluctuations of the vacuum electromagnetic field. In spite of this, stochastic quantization is an attractive alternative for all physicists felling not so comfortable with the epistemological content of the traditional interpretation of quantum mechanics.  

Recently, Haba and Kleinert \cite{haba} have pro\-po\-sed a new approach for sto\-chas\-tic quantization, hereafter referred to as the HK method. HK is based directly on the use of Newton equations with the presence of noise. After introducing an auxiliary field, defined in terms of the solutions for the stochastic dynamical equations, a deterministic partial differential equation is found averaging over all stochastic processes. This deterministic equation, after a suitable transformation, yields the Sch\"odinger equation for the mechanical system under consideration. Solving the Schr\"odinger equation reproduces the spectrum of the system. In comparison with other stochastic quantization methods like Nelson's stochastic mechanics \cite{nelson} or the Parisi and Wu method \cite{parisi, damgaard}, the HK approach seems to be remarkably concise.  Moreover, as shown in \cite{haba}, the HK method can be applied to arbitrary time-independent one-dimensional mechanical systems. 

In view of the elegance and conciseness of the HK method, it is valuable to extend it to more general dynamical  systems. Several avenues are open in this regard. For instance, one can address the generalization of HK to higher dimensionality, to composite systems, to systems exposed to external electromagnetic fields and so on. Here we consider the application of HK to a non-autonomous system, namely, the time-dependent harmonic oscillator (TDHO), characterized by a frequency function variable along time \cite{lewis}. We refrain from listing the large list of applications of the TDHO, including such several fields as quantum optics, cosmology, non-linear elasticity and hydrodynamics \cite{espinoza}. Here we limit ourselves to apply HK to this class of systems,  showing that non-autonomous systems are also amenable to stochastic quantization via the HK method. We will see that the time-dependence of the frequency can be eliminated through a rescaling transformation. 

The paper is organized as follows. In Section II, we apply the HK method to the TDHO equation. In this Section, we both present the HK method with some more details than in the original reference \cite{haba} and apply it, obtaining a stochastic differential equation simulating the TDHO. In Section III we obtain the formal solution for this stochastic differential equation and in Section IV this formal solution is used to describe the time-evolution of the mother field, one of the fundamental elements in the HK method. We then derive the Schr\"odinger equation associated to the TDHO.  Section V is devoted to the conclusions. 

\section{Stochastic Differential Equation}

Our purpose is to apply the HK method to obtain the quantization of the TDHO equation,
\begin{equation}
\ddot{x} + \omega^{2}(t)\,x = 0 \,,
\end{equation}
where $\omega(t)$ is the time-dependent frequency. The HK method consists of three steps, (a) postulate of a classical, stochastic equation, which reduces to the Newton equation of the system in the limit of zero noise; (b) introduction of an auxiliary field (the mother field) defined in terms of the solutions of the original, stochastic equation; (c) averaging over stochastic processes, yielding a deterministic equation equivalent to the Schr\"odinger equation. For the TDHO, we choose the stochastic equation
\begin{equation}
\label{e1}
\dot{\bf r} = \frac{{\bf\Omega}\times{\bf r}}{\rho^2} + \frac{\dot\rho}{\rho}\,{\bf r} + f(t)\,{\bf n} \,, 
\end{equation}
where ${\bf r} = (x,y,0)$ and ${\bf\Omega} = (0,0,\Omega)$, for constant $\Omega$, and where ${\bf n} = (1,1,0)/\sqrt{2}$ is an unitary vector in the diagonal direction in the plane. The function $\rho$ is any particular solution for the Pinney's \cite{pinney} equation, 
\begin{equation}
\label{pp}
\ddot\rho + \omega^{2}(t)\rho = \frac{\Omega^2}{\rho^3} \,,
\end{equation}
where $\omega(t)$ is the time-dependent frequency characterizing the TDHO. It is convenient to take $\Omega \neq 0$ so that $\rho$ never vanish. Finally, $f(t)$ is a stochastic variable with zero mean over statistical processes, 
\begin{equation}
\label{e2}
\left\langle f(t) \right\rangle = 0 \,,
\end{equation}
and with correlation function
\begin{equation}
\label{e3}
\left\langle f(t) f(t') \right\rangle = \hbar \, \delta(t-t') \,,
\end{equation}
so that there is a white noise stochastic process. 

The choice for the classical dynamical equation is justified because, treating formally $f$ as an ordinary function, the second order equation obtained from (\ref{e1}) is 
\begin{equation}
\ddot{\bf r} + \omega^{2}(t)\,{\bf r} = (\dot{f} + \frac{\dot\rho}{\rho}\,f)\,{\bf n} + \frac{f}{\rho^2}\,{\bf\Omega}\times{\bf n} \,,
\end{equation}
which reduces to the planar isotropic TDHO when $f = 0$. Latter on, as we shall see, the HK method proposes a dimensional reduction from 2D to 1D, so that we will recover the one-dimensional TDHO. 

To proceed, let us introduce the following rescaling, 
\begin{equation}
\label{rs}
X = x/\rho \,, \quad Y = y/\rho \,, \quad T = \int_{0}^{t}dt'/\rho^{2}(t') \,.
\end{equation}
With the new variables, we obtain from (\ref{e1}) the rescaled stochastic equation
\begin{equation}
\label{eq}
\frac{d{\bf R}}{dT} = {\bf\Omega}\times{\bf R} + F(T)\,{\bf n} \,,
\end{equation}
where ${\bf R} = (X,Y,0)$ and $F(T)$ is defined by 
\begin{equation}
F(T) = \rho(t)\,f(t) \,.
\end{equation}
Since $f$ is a stochastic function, so is $F$. The statistical properties of $F$ follows from (\ref{e2}-\ref{e3}). The new stochastic function has zero mean, 
\begin{equation}
\label{e4}
\left\langle F(T) \right\rangle = 0 \,,
\end{equation}
and has correlation function
\begin{equation}
\label{a}
\left\langle F(T) F(T') \right\rangle = \hbar\,\delta(T - T') \,,
\end{equation}
defining a white noise. 

Equation (\ref{e4}) is an immediate consequence of the zero mean of $f$. On the other hand, (\ref{a}) is demonstrated in the following way. By the definition of $F$, we have 
\begin{equation}
\label{x}
\left\langle F(T)F(T') \right\rangle = \rho(t)\rho(t') \left\langle f(t)f(t') \right\rangle  \,,
\end{equation}
where the transformed times are 
\begin{equation}
T = \int_{0}^{t}\frac{dt''}{\rho^{2}(t'')} \,, \quad T' = \int_{0}^{t'}\frac{dt''}{\rho^{2}(t'')} \,.
\end{equation}
From (\ref{x}) and the correlation function in terms of the original variables, we obtain 
\begin{equation}
\label{x1}
\left\langle F(T)F(T') \right\rangle = \hbar\rho^{2}(t)\,\delta(t-t') \,.
\end{equation}
This gives an expression for the correlation function of $F$ in terms of the original time variable. Our objective is to show that the right-hand side of (\ref{a}) coincides with this. 

It turns out that 
\begin{equation}
\label{x11}
\delta(T-T') = \delta\left(\int_{t'}^{t}\frac{dt''}{\rho^{2}(t'')}\right) \,,
\end{equation}
using the definition of rescaled time. The right-hand side of the last equation can be handled with the following property of the delta function,
\begin{equation}
\label{a2}
\delta(\varphi(t)) = \sum_{i}\frac{\delta(t-t_{i})}{|\varphi'(t_{i})|} \,, 
\end{equation}
for an arbitrary function $\varphi$ and for $\varphi(t_{i}) = 0$, $\varphi'(t_{i}) \neq 0$. The sum (\ref{a2}) is over the zeros of the function $\varphi$. Applying (\ref{a2}) to (\ref{x11}), we get
\begin{equation}
\delta(T-T') = \rho^{2}(t)\,\delta(t-t') \,, 
\end{equation}
thus showing the equivalence between (\ref{a}) and (\ref{x1}) as we desired. 

In the new variables, Equation (\ref{eq}) is the same as that used by HK in the case of the time-independent harmonic oscillator (see Eq. (3) of \cite{haba}). Hence, our remaining task is to repeat the procedure by HK and map our conclusions to the original, non-rescaled variables. We also offer some extra details on the necessary calculations. 

\section{Formal Solution}

We can obtain the formal solution for Equation (\ref{eq}) as follows. Thinking the stochastic function $F$ as an ordinary function, we easily obtain the linear invariants 
\begin{eqnarray}
\label{c1}
X_0 &=& X\cos\Omega\,T + Y\sin\Omega\,T \nonumber \\ &-&\frac{1}{\sqrt{2}}\int_{0}^{T}dT'\,F(T')(\cos\Omega\,T' + \sin\Omega\,T') \,, \\
\label{c2}
Y_0 &=& -X\sin\Omega\,T + Y\cos\Omega\,T \nonumber \\ &-&\frac{1}{\sqrt{2}}\int_{0}^{T}dT'\,F(T')(\cos\Omega\,T' - \sin\Omega\,T') 
\end{eqnarray}
for the dynamical equations. In other words, both $X_0$ and $Y_0$ are constant along trajectories and can be interpreted as the initial conditions for $X$ and $Y$ respectively. Solving (\ref{c1}-\ref{c2}) for $X$ and $Y$ we obtain 
\begin{eqnarray}
X &=& X_{0}\cos\Omega\,T - Y_{0}\sin\Omega\,T + \frac{\cos\Omega\,T}{\sqrt{2}}\int_{0}^{T}dT' F(T') (\cos\Omega\,T' + \sin\Omega\,T') \nonumber \\ &-& \frac{\sin\Omega\,T}{\sqrt{2}}\int_{0}^{T}dT' F(T') (\cos\Omega\,T' - \sin\Omega\,T') \,, \\
Y &=& Y_{0}\cos\Omega\,T + X_{0}\sin\Omega\,T + \frac{\cos\Omega\,T}{\sqrt{2}}\int_{0}^{T}dT' F(T') (\cos\Omega\,T' - \sin\Omega\,T') \nonumber \\ &+& \frac{\sin\Omega\,T}{\sqrt{2}}\int_{0}^{T}dT' F(T') (\cos\Omega\,T' + \sin\Omega\,T') \,.
\end{eqnarray}

We can use the exact solution to expand $X$ and $Y$ around $T = 0$. Integrating (\ref{a}) we get the estimate $F^{2}(\Delta T) \sim \hbar/\Delta T$ for small $\Delta T$. Therefore, $\int_{0}^{\Delta T}F(T)dT \sim \sqrt{\hbar\Delta T}$. Using this and expanding the exact solution, the result is 
\begin{eqnarray}
\label{s1}
X &=& X_0 - \Omega Y_0 \Delta T + \frac{1}{\sqrt{2}}\int_{0}^{\Delta T}dT F(T) + O((\Delta T)^{3/2}) \,, \\
\label{s2}
Y &=& Y_0 + \Omega X_0 \Delta T + \frac{1}{\sqrt{2}}\int_{0}^{\Delta T}dT F(T) + O((\Delta T)^{3/2}) \,.
\end{eqnarray}
Hence we deal with an expansion in powers of $\varepsilon = (\Delta T)^{1/2}$. More precisely,
\begin{eqnarray}
\label{ss1}
X &=& X_0 + \varepsilon\alpha_1 + \varepsilon^{2}\alpha_2 + 0(\varepsilon^3) \,,\\
\label{ss2}
Y &=& Y_0 + \varepsilon\beta_1 + \varepsilon^{2}\beta_2 + 0(\varepsilon^3)  \,,
\end{eqnarray}
where
\begin{eqnarray}
\varepsilon\alpha_1 &=& \varepsilon\beta_1 = \frac{1}{\sqrt{2}}\int_{0}^{\Delta\,T}dT F(T) \,, \\
\varepsilon^{2}\alpha_2 &=& - \Omega Y_0 \Delta T \,, \quad \varepsilon^{2}\beta_2 = \Omega X_0 \Delta T \,.
\end{eqnarray}

\section{Mother Field and Stochastic Quantization}

Following HK, the next step is the introduction of an auxiliary field, depending on the solution for the stochastic equation (\ref{eq}). Let us define a time-independent mother field ${\bf U}({\bf R}) = (U_{1}({\bf R}), U_{2}({\bf R}))$ supposed to satisfy the Cauchy-Riemann conditions
\begin{equation}
\label{cau}
\frac{\partial U_1}{\partial X} = \frac{\partial U_2}{\partial Y} \,, \quad \frac{\partial U_1}{\partial Y} = - \frac{\partial U_2}{\partial X} \,.
\end{equation}
At time $T$, we can construct a time-dependent field ${\bf U}_{T}({\bf R},T)$ which at $T = 0$ equals ${\bf U}({\bf R})$ and which is given in terms of the flow of the stochastic differential equation (\ref{eq}),
\begin{equation}
{\bf U}_{T}({\bf R},T) = {\bf U}({\bf R}(T)) \,,
\end{equation}
where ${\bf R}(T)$ evolves in time according to (\ref{eq}). Using (\ref{ss1}-\ref{ss2}), we can expand the time-dependent field around $T = 0$ to obtain
\begin{eqnarray}
{\bf U}_{T}({\bf R}, \Delta\,T) \! = \!{\bf U}({\bf R})\!\! &+& \!\!\Delta T\,({\bf\Omega}\times\,{\bf R}\cdot\bar{\nabla}){\bf U}({\bf R}) + \int_{0}^{\Delta T} dT F(T) ({\bf n}\cdot\bar{\nabla}){\bf U}({\bf R}) \nonumber \\ 
&+& \frac{1}{2}\int_{0}^{\Delta T}dT\int_{0}^{\Delta T}dT' F(T) F(T')({\bf n}\cdot\bar{\nabla})^{2}\,{\bf U}({\bf R}) \nonumber \\ \label{exp} &+& 0((\Delta T)^{3/2}) \,.
\end{eqnarray}
where $\bar{\nabla}$ is the gradient operator in ${\bf R}$ coordinates.

By construction, ${\bf U}_{T}({\bf R},T)$ is a stochastic function. However, an ordinary function $\bar{\bf U}({\bf R},T)$ can be build after performing the last step in the HK method, i.e., averaging over stochastic processes, 
\begin{equation}
\bar{\bf U}({\bf R},T) = \left\langle {\bf U}_{T}({\bf R},T) \right\rangle \,.
\end{equation}
Averaging (\ref{exp}) and taking into account the statistics of $F$, we obtain
\begin{equation}
\label{h}
\frac{\partial\bar{\bf U}}{\partial T} = \hat{H}\,\bar{\bf U} \,,
\end{equation}
with the time evolution operator
\begin{equation}
\label{hh}
\hat{H} = \left(({\bf\Omega}\times{\bf R})\cdot\bar{\nabla}\right) + \frac{\hbar}{2}\left({\bf n}\cdot\bar{\nabla}\right)^2 \,.
\end{equation}
As shown by HK, the averaged mother field is harmonic for all times. 

Equation (\ref{h}) is equivalent to a Schr\"odinger equation after restricting to the $Y = 0$ line. To show this, notice that the averaging process preserves the harmonic property (\ref{cau}). Using the Cauchy-Riemann properties, we can rewrite (\ref{h}) as
\begin{eqnarray}
\frac{\partial\bar{U_1}}{\partial T} &=& - \Omega\,\left(X\frac{\partial \bar{U}_2}{\partial X} + Y\frac{\partial \bar{U}_1}{\partial X}\right) - \frac{\hbar}{2}\frac{\partial^{2}\bar{U}_2}{\partial X^2}            \,, \\
\frac{\partial\bar{U_2}}{\partial T} &=&  \Omega\,\left(X\frac{\partial \bar{U}_1}{\partial X} - Y\frac{\partial \bar{U}_2}{\partial X}\right) + \frac{\hbar}{2}\frac{\partial^{2}\bar{U}_1}{\partial X^2}   \,.
\end{eqnarray}
Restricting to $Y = 0$ and defining the complex field 
\begin{equation}
\bar{\psi}(X,T) = \exp\left(-\frac{\omega\,X^2}{2\hbar} - \frac{i\omega\,T}{2}\right)\,\left(\bar{U}_{1}(X,0,T) + i\bar{U}_{2}(X,0,T)\right) \,,
\end{equation}
we obtain the Schr\"odinger equation for a one-dimensional time-independent harmonic oscillator, 
\begin{equation}
\label{ti}
i\hbar\frac{\partial\bar{\psi}}{\partial T} = \left(-\frac{\hbar^2}{2}\frac{\partial^2}{\partial X^2} + \frac{\Omega^{2}X^2}{2}\right)\bar{\psi} \,,
\end{equation}
with frequency $\Omega$. 

To obtain the quantization of the TDHO, consider the transformation 
\begin{equation}
\psi = \rho^{-1/2} \exp\left(\frac{i\dot\rho\,x^2}{2\hbar\rho}\right)\,\bar{\psi} \,,
\end{equation}
where $\rho$ is a solution for the Pinney equation (\ref{pp}). Using the Pinney equation and (\ref{ti}) and inverting the rescaling (\ref{rs}), we get 
\begin{equation}
\label{td}
i\hbar\frac{\partial\psi}{\partial t} = \left(-\frac{\hbar^2}{2}\frac{\partial^2}{\partial x^2} + \frac{\omega^{2}(t)x^2}{2}\right)\,\psi \,.
\end{equation}
The spectrum of the quantum one-dimensional TDHO (\ref{td}) can be obtained, for instance, solving the Pinney equation and applying the Lewis-Riesenfeld method \cite{lewisr}. 

\section{Conclusion}

We have obtained the quantization of the TDHO using the HK method. Our approach relies on a rescaling transformation, which removes the explicit time-dependence from the classical stochastic Newtonian equation. We may observe that, since the quantization procedure depends on the function $\rho(t)$, which is any arbitrary solution for the Pinney equation, we have in fact an infinite family of quantizations. 

Since, essentially, our procedure was based on a rescaling transformation, it remains the question of what other classes of explicitly time-dependent mechanical systems are amenable to the HK method. In addition, we have not touched upon the measurement theory associated to HK or the physical origin of the noise. Finally, in the case of HK there are other possible and necessary extensions, like consideration of higher dimensional cases, many particle systems and the inclusion of general electromagnetic fields. With the treatment of many particle systems we can address another fundamental property of quantum mechanics, namely, entanglement.

\end{document}